\documentclass[aps,twoside,twocolumn]{revtex4}
\usepackage{amssymb}
\usepackage{amsmath}
\usepackage{graphicx}

\newcommand{\be}{\begin{eqnarray}}
\newcommand{\ee}{\end{eqnarray}}

\begin{document}
\title[]{Photon Number Splitting of Squeezed Light by a Single Qubit in Circuit QED}

\author{Kyungsun \surname{Moon}}
\affiliation{Department of Physics, Yonsei University, Seoul 120-749, Korea }
\date{\today}

\begin{abstract}

We theoretically propose an efficient way to generate and detect squeezed light by a single qubit in circuit QED.
By tuning the qubit energy splitting close to the fundamental frequency of the first harmonic mode (FHM) in a transmission line resonator and placing the qubit at the nodal point of the third harmonic mode, one can generate the resonantly enhanced squeezing of the FHM upon pumping with the second harmonic mode.
In order to investigate the photon number splitting for the squeezed FHM, we have numerically calculated the qubit absorption spectrum, which exhibits regularly spaced peaks at frequencies separated by twice the effective dispersive shift. It is also shown that adding a small pump field for the FHM makes additional peaks develop in between the dominant ones as well.

\end{abstract}



\maketitle

Recent experimental developments in producing strongly squeezed light have made the ultra-precision quantum measurement possible approaching to the sensitivity limit of the gravitational wave detection\cite{LIGO}. Spatially separated entangled microwave photon pairs have been successfully generated by the superconducting on-chip parametric amplifier, which can be potentially useful for quantum communications\cite{Devoret}. Using the Josephson parametric amplifier, an efficient two-mode squeezing in the microwave domain has also been experimentally achieved\cite{Wallraff,Yurke}.
At the atomic scale, an interesting experiment to generate squeezed light from a single atom has been performed using high-finesse optical cavity and achieved a weak but noticeable squeezing on the order of $10 {\rm mdB}$\cite{SingleAtomSqueezing}. Following the remarkable progress in the cavity QED, the circuit QED has been very successful in experimentally realizing several outstanding issues in quantum optics such as resolving the photon number states\cite{Haroche,NumberSplitting1,NumberSplitting2}.

In the paper, we theoretically study the squeezing of the microwave mode in a transmission line resonator by a single qubit. Due to the strong coupling between microwave modes and a qubit in circuit QED\cite{circuitQED,transmon}, the qubit can act as an optical coupler and generate several important non-linear optical processes. Among them, we will mainly focus on the following two processes, which will become dominant upon strong pumping with the second harmonic microwave field $E_2$ with frequency $\omega_2$. Here we will tune the qubit energy splitting $E_{\rm 01}$ close to the fundamental frequency $\omega_1$ of the FHM in transmission line resonator and $\omega_2=2\omega_1$.
The first process describes the resonantly enhanced squeezing of the FHM, which can be represented by $a_1^\dagger a_1^\dagger a_2$. The operators $a_n (a_n^\dagger)$ denote the annihilation(creation) operator for the $n-th$ harmonic mode. It demonstrates degenerate parametric down-conversion, where an incident photon with frequency $\omega_2$ is down-converted into a pair of photons with frequency $\omega_1$\cite{MoonGirvin}.
The other process describes a competing process against squeezing, which also becomes resonantly enhanced upon strong pumping with the second harmonic mode and can be represented by $a_1^\dagger a_2^\dagger a_3$\cite{Jaehne}. While the degenerate parametric down-conversion creates photon pairs with frequency $\omega_1$, this term creates a single photon with frequency $\omega_1$. By creating unpaired photon, it will strongly degrade the quality of squeezing\cite{SingleAtomSqueezing}.
Since these two terms arise and become enhanced simultaneously, it is quite difficult to suppress one over the other.
For this purpose, the circuit QED has a great advantage over the cavity QED. Unlikely from the cavity QED, the coupling strengths between the qubit and the individual microwave harmonic modes can be readily controlled by varying the position of the qubit relative to the transmission line resonator in circuit QED\cite{Blais}. Here we will place the qubit at the nodal position of the third harmonic mode, which will strongly reduce the coupling strength between the third harmonic mode and the qubit. Hence the unwanted competing channel can be effectively blocked leading to the possibility of generating the strong squeezed light by single qubit in circuit QED.

For the appropriate choice of experimentally relevant set of parameters\cite{NumberSplitting1,NumberSplitting2},
we have first calculated the Homodyne spectrum $S_a(\omega)$ of the propagating output microwave field as a function of frequency detuning $\Delta\omega=\omega-\omega_1$, which has served as a standard physical quantity for detecting squeezed light in quantum optics experiment\cite{WallMilburn}. It has been shown that for a given pump field $E_2$ of the second harmonic mode, $S_a(\omega)$ exhibits a Lorentzian dip at the characteristic frequency detuning $\Delta\omega=\bar{\chi}$ corresponding to the effective dispersive shift in the presence of squeezing. As the pump field $E_2$ of the second harmonic mode increases, the minimum value of Homodyne spectrum $S_{min}$ at the dip decreases and approaches close to zero just below the critical pump field. It will eventually reach to the minimum value set both by the finite dissipations of the system and by the total photon number $N$ used in our numerical simulations.
We have also numerically calculated the qubit absorption spectrum $S_\sigma(\omega)$ as a function of frequency detuning $\Delta\omega=\omega-E_{\rm 01}$, which remarkably exhibits a regularly spaced peaks at frequencies separated by twice the effective dispersive shift. Upon applying a small coherent pump for the FHM, it has been demonstrated that additional peaks appear at frequencies in between the dominant ones heralding the emergence of the single photon processes. Hence the experimental observation of photon number splitting can be used to detect the squeezed light in our circuit QED setup.

We will start with the following Hamiltonian for circuit QED in the dispersive regime of $g_1<<|\Delta|<<\omega_1$ at a given gate charge $N_g$ (Details of derivation are given in Appendix A)
\begin{eqnarray}
H&=&{E_{\rm 01}\over 2}\sigma_z + \sum_n\omega_n a_n^\dagger a_n+{(g_1\sin\theta)^2\over 2\Delta}(2 a_1^\dagger a_1+1)\sigma_z \nonumber\\
&+&\sum_n{1\over{n(n+1)}}{{g_1 g_n g_{n+1}}\over \Delta\omega_1}\sin\theta\sin 2\theta\nonumber\\
&\times&(a_1^\dagger a_n^\dagger a_{n+1}+a_1 a_n a_{n+1}^\dagger)\sigma_z\nonumber\\
&+&\sum_{n=1,2}(E_n(t)a_n e^{i\omega_n t}+E_n^*(t)a_n^\dagger e^{-i\omega_n t}),
\label{Hamiltonian2}
\end{eqnarray}
where $\sigma_z=\pm 1$ represent two qubit states and $g_n$ denotes the coupling strength between qubit and the $n-th$ harmonic mode with frequency $\omega_n$. $\Delta=E_{\rm 01}-\omega_1$ represents the energy detuning. For the charge qubit with $E_J/E_C << 1$, the angle $\theta=\tan^{-1} (E_J/E_{el})$, where $E_J$ represents the Josephson energy of a qubit and $E_{el}=4E_C (1-2N_g)$ with $E_C$ the charging energy of a qubit and $N_g$ the gate charge, which can be tunable by varying the gate voltage. For the transmon qubit with $E_J/E_C >> 1$,
the appropriate Hamiltonian has been derived in Appendix B.


The first two terms in Eq. \ref{Hamiltonian2} represent the Hamiltonian for the qubit with $E_{\rm 01}$ being the qubit energy splitting and that for the cavity photons with discrete angular frequencies $\omega_n=n\omega_1$ respectively. The third term represents the cavity pull of the $\omega_1$ photon or the ac-Stark shift of the qubit. The fourth terms have been derived from the third order processes, which represent various nonlinear optical processes and the contributions of these terms will be quite small in general.
Here we will mainly focus on the squeezing term given by $(g_1^2 g_2/ 2\Delta\omega_1)\sin\theta\sin 2\theta a_1^\dagger a_1^\dagger a_2\sigma_z$,
where an incident photon with frequency $\omega_2$ into the cavity is down-converted to a pair of photons with frequency $\omega_1$. The coupling constants $g_n$ are given by the following formula: $g_n=(eC_g/C_\Sigma)\sqrt{\hbar\omega_n/Lc}\sin(n\pi x/L)$ for odd $n$ and $(eC_g/C_\Sigma)\sqrt{\hbar\omega_n/Lc} \cos(n\pi x/L)$ for even $n$, where $x$ represents the position of the qubit with $-L/2\le x \le L/2$ and $L$ the length of the cavity resonator. Here $C_g, C_\Sigma$, and $c$ represent various capacitances in our circuit QED system\cite{Blais}.
The last term stands for the driving microwave field at the following two frequencies of $\omega_1$ and $\omega_2$.

In order to generate an intense squeezing signal, it will require a strong pumping of the microwave field for the second harmonic mode with frequency $\omega_2$. Since the second harmonic mode is off-resonant from the qubit, the cavity pull for the second harmonic mode is negligible and hence the strong pumping will be readily achievable experimentally.

Apart from the squeezing term, there exists an unwanted competing term against squeezing given by $(g_1 g_2 g_3/6\Delta\omega_1)\sin\theta\sin 2\theta a_1^\dagger a_2^\dagger a_3\sigma_z$, which also becomes strongly enhanced upon strong pumping with the second harmonic mode. While the squeezing term creates a pair of photons with frequency $\omega_1$, this term creates or annihilates a single photon with frequency $\omega_1$ and hence breaks a pair, which will strongly degrade the quality of squeezing.
In circuit QED, the coupling strengths between the qubit and the microwave harmonic modes can be controlled by varying the position of the qubit, which can help overcome this difficulty.
Here we will place the qubit at the nodal position $x=L/3$ of the third harmonic mode, which will strongly suppress the coupling between the third harmonic mode and the qubit and hence effectively shut off the competing channel.

In the strong pump limit, one can take the quantum annihilation operator $a_2$ of the second harmonic mode to be classical with a value of $a_2=(2 E_2/\kappa_2) e^{-i\omega_2 t}$, where $\kappa_2$ represents the cavity loss for the second harmonic photon. In the rotating frame of the qubit at frequency $E_{\rm 01}$ and of the photon at frequency $\omega_1$, one can obtain the following effective Hamiltonian $H_{\rm eff}$
\begin{equation}
H_{\rm eff}=\chi(a^\dagger a+{1\over 2})\sigma_z + \chi_{sq}(a^\dagger a^\dagger+ a a)\sigma_z + E_1 (a^\dagger+a),
\label{Hamiltonian3}
\end{equation}
where we have omitted the indices for $a_1, a_1^\dagger$, $\chi=(g_1\sin\theta)^2/ \Delta$, and $\chi_{sq}=(E_2/\kappa_2)(g_1^2 g_2/
\Delta\omega_1)\sin\theta\sin 2\theta$.
The last term in Eq. \ref{Hamiltonian3} describes the pump field for the FHM, which will displace $a$ and $a^\dagger$ by $E_1\sigma_z/(\chi+2\chi_{sq})$.
The magnitude of squeezing term $\chi_{sq}$ is proportional to the pump amplitude $E_2$.

Based on the above Hamiltonian, we will investigate the characteristics for squeezing by a single qubit in circuit QED.
We will introduce the transformed photon operators $b$ and $b^\dagger$, which represent the annihilation and creation operators for two-photon coherent state, through the following unitary transformations\cite{WallMilburn}: $b=S^\dagger (-r) a S(-r)$ and $b^\dagger=S^\dagger (-r) a^\dagger S(-r)$ with $S(-r)=\exp[r(a^\dagger a^\dagger - a a)/2]$. By choosing the parameter $r$ to satisfy the following relation $\tanh 2r=2\chi_{sq}/\chi$, one can diagonalize the above Hamiltonian $H_{\rm eff}$ for $\chi_{sq}<\chi/2$, which is given by
\begin{equation}
H_{\rm eff}=\bar{\chi}(b^\dagger b + {1\over 2})\sigma_z
+ E_1 \left( \frac {\chi-2\chi_{sq}}{\chi+2\chi_{sq}}\right)^{1/4}(b^\dagger + b),
\label{Hamiltonian4}
\end{equation}
where $\bar{\chi}=\sqrt{\chi^2-(2\chi_{sq})^2}$ represents the effective dispersive shift in the presence of squeezing.
In the absence of a coherent pump $E_1=0$ for the FHM, the energy eigenstates of the system are the photon number states for ${\hat N}_b=b^\dagger b$, which are composed of squeezed vacuum state $\vert 0\rangle_{sq}$ and the excited states $\vert n_b\rangle_{sq}$. It is quite interesting to notice that the qubit is directly coupled to the number operator ${\hat N}_b$ and the coupling constant $\chi$ is replaced with the effective dispersive shift $\bar{\chi}$. For $\chi_{sq}=0$, one can reproduce the ac-Stark shift term proportional to ${\hat N}_a=a^\dagger a$.
Here the ideal squeezed vacuum state is given by $\vert 0\rangle_{sq}=S(r)\vert 0\rangle$, which satisfies that $b|0\rangle_{sq}=0$. It is well known that for the ideal squeezed vacuum state, the variance of a squeezed quadrature $X_1=a+a^\dagger$ is reduced below that of a coherent state such that $(\Delta X_1)^2=e^{-2r}=[(\chi-2\chi_{sq})/(\chi+2\chi_{sq})]^{1/2}$.
For the conjugate quadrature $X_2=-i(a-a^\dagger)$, the variance is amplified such that $(\Delta X_2)^2=[(\chi+2\chi_{sq})/(\chi-2\chi_{sq})]^{1/2}$ satisfying the minimum uncertainty condition\cite{Scully}.

Now we will carry out a detailed numerical analysis for realistic squeezing based on the Hamiltonian $H_{\rm eff}$ of Eq. \ref{Hamiltonian3} in the finite Hilbert space with total photon number $N-1$. We have taken into account the finite cavity loss $\kappa$, qubit decay rate $\gamma$ and pure dephasing rate $\gamma_\phi$ by adding the Lindblad type dissipation terms. The time evolution of the density matrix $\rho$ for the system can be described by the following equation
\begin{equation}
{d\rho\over dt}=-{i\over \hbar}\left[H,\rho\right] + \kappa {\cal D}[a]\rho + \gamma_1 {\cal D}[\sigma_-]\rho +{\gamma_\phi\over 2}  {\cal D}[\sigma_z]\rho,
\label{MasterEquation}
\end{equation}
where the Lindblad superoperator is given by ${\cal D}[L]\rho=\left( 2L\rho L^\dagger -L^\dagger L \rho-\rho L^\dagger L\right)/2$.

\begin{figure}
\includegraphics[width=3.5in,height=1.6in]{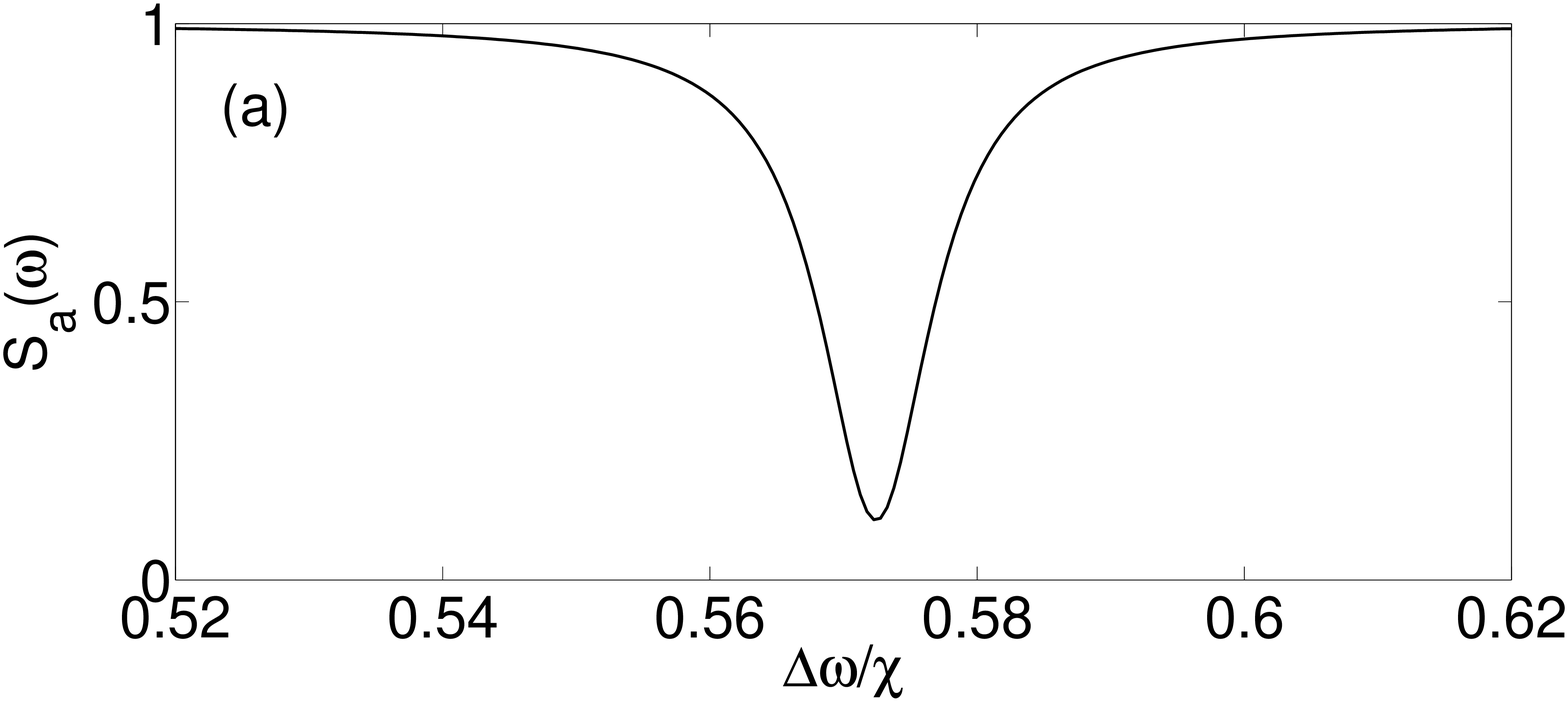}
\includegraphics[width=3.5in,height=1.6in]{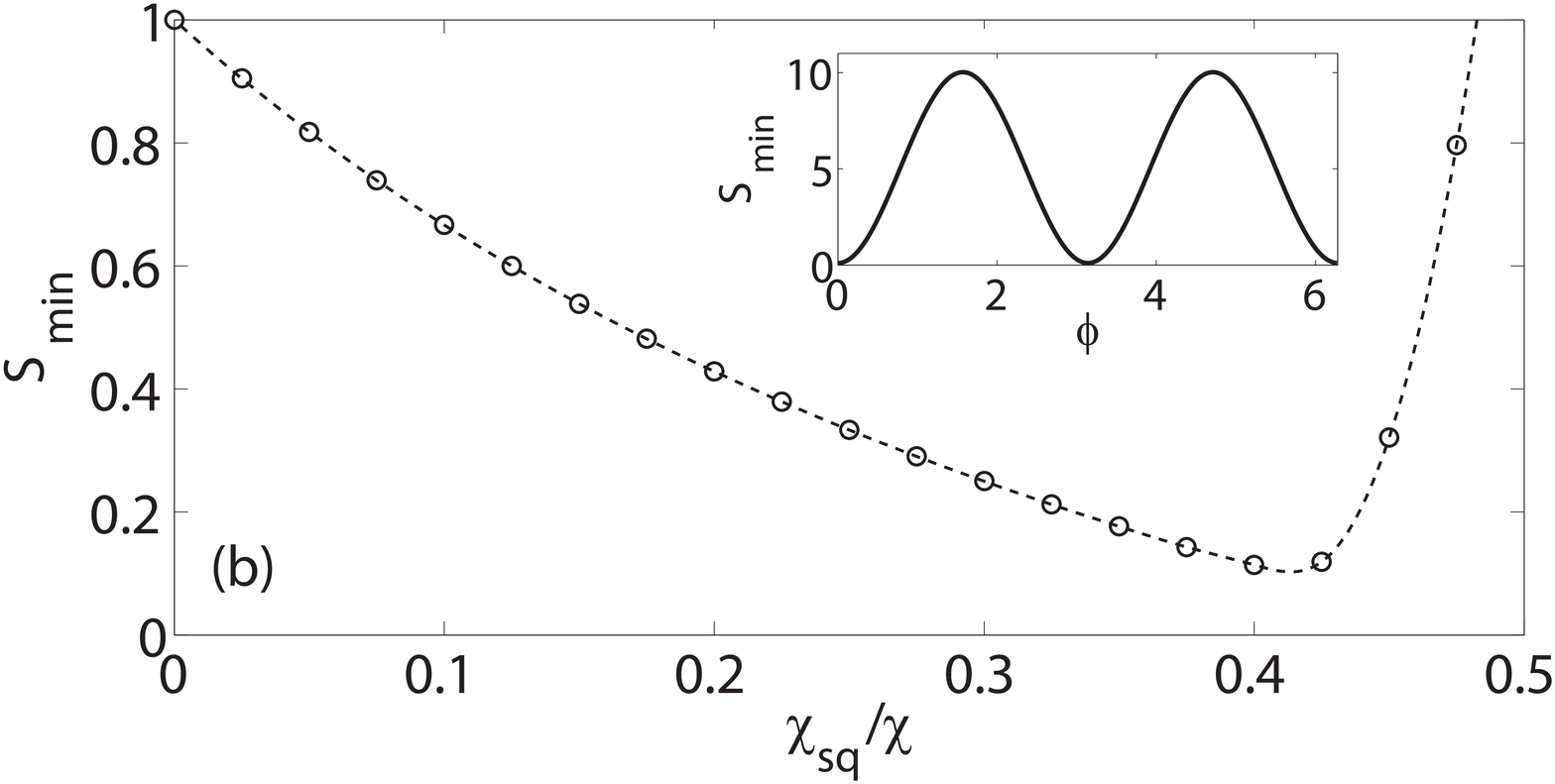}
\caption{ (a) The Homodyne spectrum $S_a(\omega)$ is plotted as a function of frequency detuning $\Delta\omega$ for a given value of $\chi_{sq}=0.41$ exhibiting the maximum squeezing at around $\Delta\omega=\bar{\chi}=\sqrt{\chi^2-(2\chi_{sq})^2}\cong 0.57$. Here $\Delta\omega=\omega-\omega_1$ and $\chi_{sq}$ are given in units of $\chi$.
(b) The maximum value of squeezing $S_{min}$ is plotted as a function of $\chi_{sq}$. In the inset, $S_{min}$ is plotted as a function of the local oscillator angle $\phi$ for $\chi_{sq}=0.41$. }
\label{Figure1}
\end{figure}

We have chosen the following experimentally relevant set of parameters for the calculation: $\kappa=0.01, \gamma_1=0.01$, and $\gamma_\phi=0.067$ in units of $\chi$ with $\chi=1$\cite{NumberSplitting1,NumberSplitting2}.
We will first consider squeezing of the propagating output microwave field with highly asymmetric cavity losses, which can be experimentally measurable by the standard Homodyne detection\cite{WallMilburn}. Following the well-established input-output formalism, one can calculate
the Homodyne spectrum $S_a(\omega)$ for the quadrature $X=e^{i\phi}a+e^{-i\phi}a^\dagger$ at a given local oscillator angle $\phi$ using the following formula
\begin{eqnarray}
S_a(\omega)&=&1+\kappa\int\limits_{-\infty}^{\infty} d\tau e^{-i\omega \tau}
{\rm Tr}\large\{ (e^{i\phi}a+e^{-i\phi}a^\dagger)\nonumber\\
&\times& e^{{\cal L}\tau}\left( e^{i\phi}a\rho_{ss}+e^{-i\phi}\rho_{ss}a^\dagger\right)\large\},
\label{HomodyneSpectrum}
\end{eqnarray}
where ${\cal L}$ denotes the Liouvillian superoperator, $\rho_{ss}$ the steady state density matrix. We have numerically calculated $S_a(\omega)$ by varying $\chi_{sq}$ and the angle $\phi$ for $N=30$. In Fig. 1(a), $S_a(\omega)$ is plotted as a function of frequency detuning $\Delta\omega=\omega-\omega_1$ for a given value of $\chi_{sq}=0.41$, which exhibits the maximum squeezing at around the effective dispersive shift $\bar{\chi}=\sqrt{\chi^2-(2\chi_{sq})^2}\cong 0.57$. In Fig. 1(b), we have plotted $S_{min}$ as a function of $\chi_{sq}$. One can see that as $\chi_{sq}$ increases to $\chi/2$, $S_{min}$ decreases reaching to the minimum value and then turns upward.
For a fixed value of $\chi_{sq}=0.41$, we have plotted $S_{min}$ as a function of the local oscillator angle $\phi$, which clearly demonstrates a strong amplification of the variance for the conjugate quadrature $X_2$ with $\phi=\pi/2$ as shown in the inset of Fig. 1(b).
We have explicitly shown that adding a finite coherent pump $E_1$ for the FHM does not change $S_a(\omega)$, since it only shifts the origin of $a$ and $a^\dagger$ operators and the qubit remains to stay in the ground state at the steady state limit.
In order to estimate the limitation of squeezing due to finite dissipations and cavity loss, we have performed the numerical simulations based on the Hamiltonian of \ref{Hamiltonian1B} and obtained $S_{min}\cong 0.06$.

\begin{figure}
\includegraphics[width=3.5in,height=1.6in]{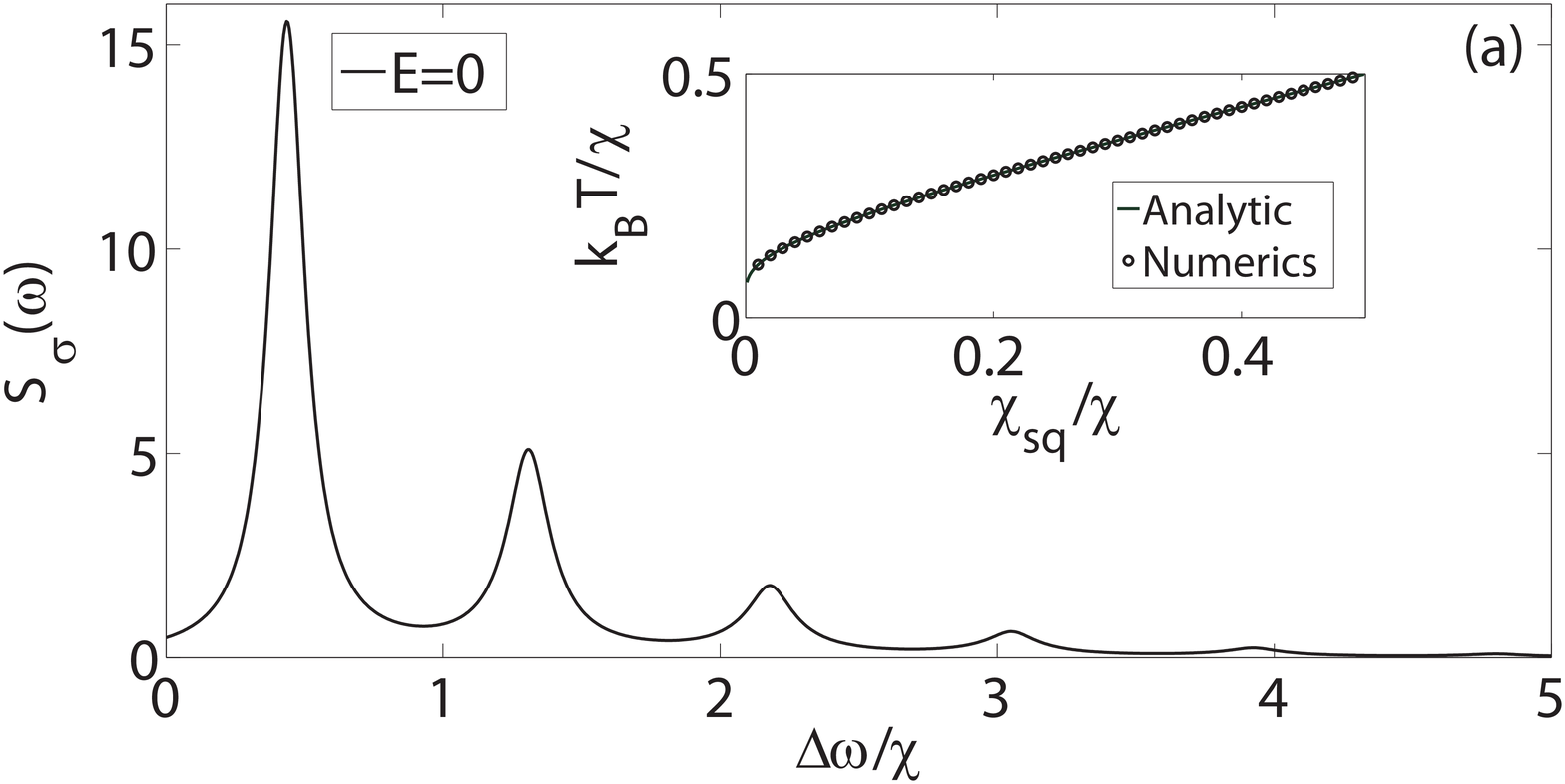}
\includegraphics[width=3.5in,height=1.6in]{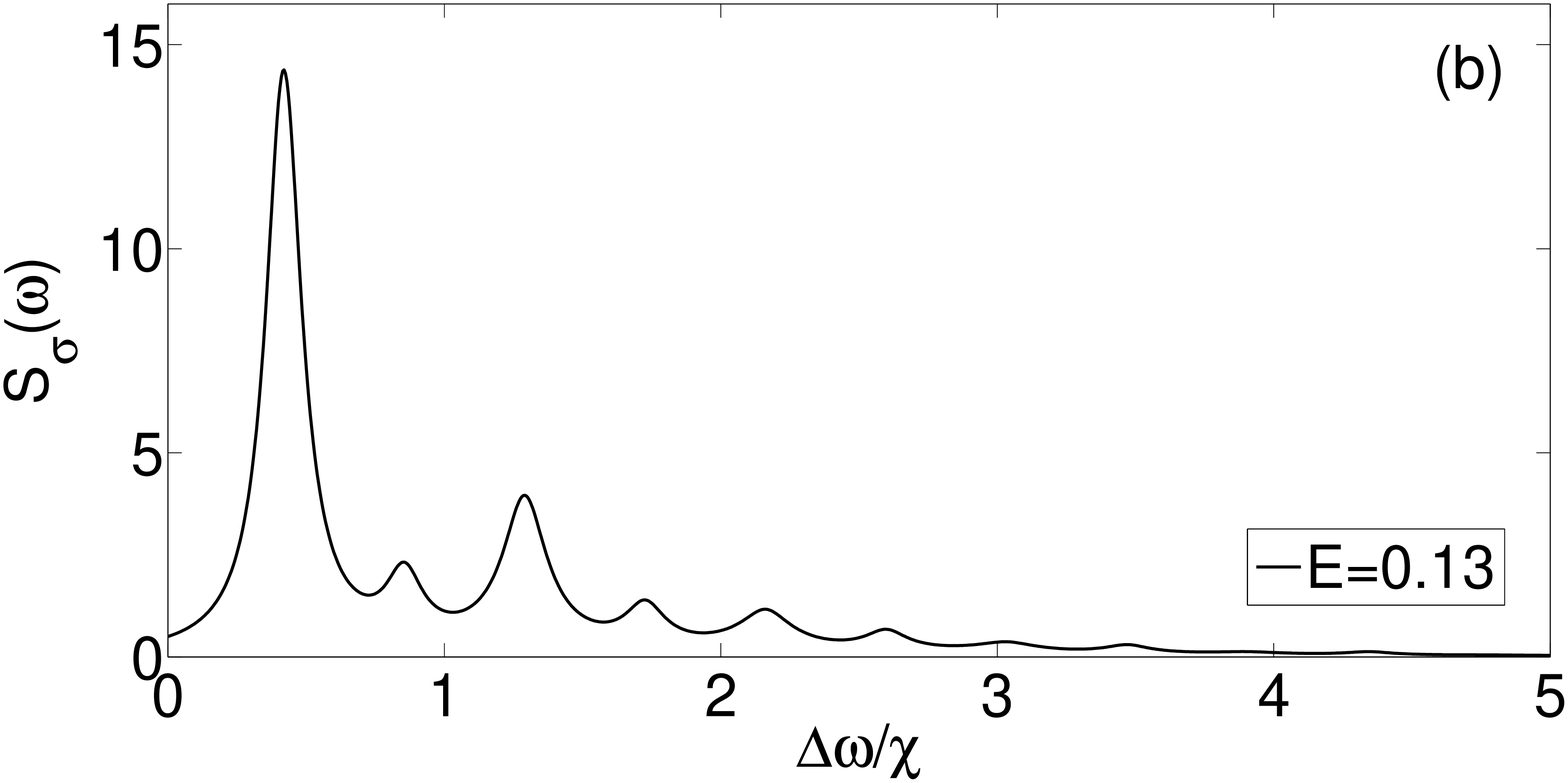}
\caption{(a) The qubit absorption spectrum $S_\sigma(\omega)$ is plotted as a function of frequency detuning $\Delta\omega$ in the absence of the coherent pump $E_1=0$ of the FHM. Here $\Delta\omega=\omega-E_{\rm 01}$, $\chi_{sq}$, and $E_1$ are given in units of $\chi$.
In the inset, $k_B T/\chi$ is plotted as a function of $\chi_{sq}/\chi$. (b) $S_\sigma(\omega)$ is plotted as a function of frequency detuning $\Delta\omega$ for $\chi_{sq}=0.45$ for a small coherent pump $E_1=0.13$ for the FHM.}
\label{Figure2}
\end{figure}

Finally we have calculated the qubit absorption spectrum $S_\sigma(\omega)$ given by
\begin{equation}
S_\sigma(\omega)={1\over {2\pi}}\int\limits_{-\infty}^{\infty} d t e^{i\omega t}
\langle \sigma_{-}(t)\sigma_{+}(0)\rangle_s,
\label{QubitSpectralFunction}
\end{equation}
where $\langle\cdots\rangle_s$ means that the average is taken over the steady state.
For a fixed value of $\chi_{sq}=0.45$, we have calculated the spectral function $S_\sigma(\omega)$ of the qubit with $E_1=0$ and $0.13$ for $N=40$.
In Fig. 2(a), $S_\sigma(\omega)$ is plotted as a function of frequency detuning $\Delta\omega=\omega-E_{\rm 01}$ in the absence of coherent pump for the FHM $(E_1=0)$, which exhibits regularly spaced peaks separated by $2\bar{\chi}\cong 0.87$, that is, twice the effective dispersive shift. It has been theoretically suggested that in the absence of squeezing term, the pump field $E_1$ for the FHM can generate a coherent state and $S_\sigma(\omega)$ will exhibit the regularly spaced peaks separated by $\chi$, which correspond to the discrete photon number states\cite{NumberSplitting1}. Subsequently the photon number splitting in the circuit QED has been experimentally resolved\cite{NumberSplitting2}.
In the presence of squeezing, $\chi_{sq}$ acts as a pump field for a photon pair $a^\dagger a^\dagger$ in analogy to the $E_1$ field for a single photon. We have also noticed that the coupling of the qubit state to the photon number operator ${\hat N}_a$ has been converted to that of the qubit state to ${\hat N}_b$. Hence there exists an interesting analogy between the original photon number states and the two-photon coherent states.

Based on the Hamiltonian of Eq. \ref{Hamiltonian4}, one can write the qubit absorption spectrum $S_\sigma(\omega)$ for small dissipations as follows
\begin{eqnarray}
S_\sigma(\omega)&=&{1\over {2\pi}}\int\limits_{-\infty}^{\infty} d t e^{i\omega t}{\rm Tr}\left\{e^{-i(2{\hat N}_b+1)\bar{\chi}t}\rho_{ss}\right\}\nonumber\\
&=&\sum_{n_b}P(n_b)\delta (\Delta\omega-(2n_b+1)\bar{\chi}),
\label{QubitSpectralFunction}
\end{eqnarray}
where $P(n_b)$ represents the number distribution function for the $n_b-th$ excited two-photon coherent state.
Hence $S_\sigma(\omega)$ will exhibit regularly spaced peaks at frequency detunings of $\bar{\chi}(2n_b+1)$, which are separated by $2\bar{\chi}$ instead of $\chi$ as clearly demonstrated in Fig. 2(a).

One can also understand the peak separations by $2\bar{\chi}$ in terms of the original $a$ photons.
Since all the dynamical processes will involve a photon pair instead of a single photon, the spectrum will
be consisted of peaks separated by twice the characteristic frequency of the system.
Now we want to calculate the number distribution function $P(n_b)$ for the steady state. By transforming to the two-photon coherent state basis, we have performed a numerical
calculation by varying $\chi_{sq}$ to obtain the diagonal component of $\rho_{ss}$, which is the number distribution function: $P(n)=(\rho_{ss})_{nn}$. From the numerical result, we have demonstrated that $P(n)$ can be very accurately described by the thermal population of the form $e^{-\hbar\bar{\chi}n/k_B T}/(1-e^{-\hbar\bar{\chi}/k_B T})$.
The effective temperature $T$ can be analytically obtained as follows.
By transforming to the two-photon coherent state basis, the Lindblad dissipation term for cavity loss generates the following two terms related to the 'thermal' photon populations: $\kappa \cosh^2 r {\cal D}[b]\rho+\kappa \sinh^2 r {\cal D}[b^\dagger]\rho$.
Since $\cosh^2 r=n_{th}+1$ and $\sinh^2 r=n_{th}$ with $n_{th}$ being the average photon number at $T$,
one can define the effective temperature $T$ as follows: $k_B T=\hbar\bar{\chi}/\ln[(\chi+\bar{\chi})/(\chi-\bar{\chi})]$.
In the inset of Fig. 2(a), $k_B T/\chi$ is plotted as a function of $\chi_{sq}/\chi$. The solid line represents the curve obtained from the analytical formula and the open circles the numerical data. One can see the excellent agreement between the analytical formula and the numerical result.
We also note that as $\chi_{sq}$ goes to zero, $T$ goes to zero very slowly as $1/\ln[\infty]$.

We now add a small coherent pump $E_1$ for the FHM.
As shown in the Hamiltonian of Eq. \ref{Hamiltonian3}, the finite $E_1$ acts as a coherent pump field for the original photon and can generate an unpaired single photon. Due to the emergence of dynamical processes involving a single photon, one can naturally expect to observe additional peaks developed in between the dominant peaks.
In Fig. 2(b), the numerical result for $S_\sigma(\omega)$ is plotted as a function of frequency detuning $\Delta\omega$ for $\chi_{sq}=0.45, E_1=0.13$, and $N=40$.
It is clearly demonstrated that the additional peaks appear at frequency detunings of $\bar{\chi}(n_b+1)$ in between the dominant ones as expected.

In summary, we have shown that by placing a qubit at the nodal point of the third harmonic mode, one can generate a squeezed light by a single qubit in circuit QED for the experimentally relevant set of parameters. It is quite remarkable to notice that the qubit absorption spectrum exhibits regularly spaced peaks at frequencies separated by twice the effective dispersive shift in the presence of squeezing. We propose to measure the photon number splitting experimentally to detect the squeezed light in our circuit QED setup.

\begin{acknowledgments}
 We want to thank the Korea Institute for Advanced Study for their hospitality, where this work has been partly done. This research was supported by Basic Science Research Program through the National Research Foundation of Korea(NRF) funded by the Ministry of Education, Science and Technology(NRF-2012R1A1A2006927).
\end{acknowledgments}

\begin{appendix}

\section{}
We will derive the effective Hamiltonian of Eq. \ref{Hamiltonian2} based on the well-established Hamiltonian describing circuit QED at a given gate charge $N_g$\cite{Blais}
\begin{eqnarray}
H&=&{E_{\rm 01}\over 2}\sigma_z + \sum_n\omega_n a_n^\dagger a_n-\sum_n g_n (a_n+a_n^\dagger)\nonumber\\
&\times& (1-2N_g-\cos\theta \sigma_z)-\sum_n g_n\sin\theta(a_n+a_n^\dagger)\sigma_x \nonumber\\
&+&\sum_{n=1,2}(E_n(t)a_n e^{i\omega_n t}+E_n^*(t)a_n^\dagger e^{-i\omega_n t}),
\label{Hamiltonian0}
\end{eqnarray}
where the angle $\theta=\tan^{-1} (E_J/E_{el})$ and $E_{el}=4E_C (1-2N_g)$.
The first two terms in Eq. \ref{Hamiltonian0} represent the Hamiltonian for the qubit and the cavity photons with discrete angular frequencies $\omega_n=n\omega_1$ respectively. The third term induces a linear displacement of the position operator $x_n=a_n+a_n^\dagger$ of the cavity photon by
$-(2g_n/\omega_n)(1-2N_g-\cos\theta \sigma_z)$.
One can see that at the charge degeneracy point(CDP) of $N_g=1/2$, the angle $\theta$ is equal to $\pi/2$ and hence the the third term vanishes.
The fourth term represents the strong coupling between the qubit and the cavity photons, which will be reduced to the Jaynes-Cummings Hamiltonian exactly at the CDP upon applying the rotating wave approximation. The last term stands for the driving microwave field at the following two frequencies of $\omega_1$ and $\omega_2$. By applying the following unitary transformation $H_1=U_1 H U_1^\dagger$ with $U_1=\exp[\sum_n (g_n/\omega_n) (a_n-a_n^\dagger)(1-2N_g-\cos\theta \sigma_z)]$, one can make the linear coupling term to disappear. Here we will focus on the dispersive regime of $g_1<<|\Delta|<<\omega_1$. In this limit, the Hamiltonian can be written as follows up to the order of ${\cal{O}}(g_n^2 / \omega_n)$
\begin{eqnarray}
H_1&=&{E_{\rm 01}\over 2}\sigma_z + \sum_n\omega_n a_n^\dagger a_n-g_1\sin\theta(a_1\sigma^\dagger+a_1^\dagger\sigma^{-})\nonumber\\
&-&\sum_n {1\over{n(n+1)}}{{g_n g_{n+1}}\over \omega_1}\sin 2\theta(a_n^\dagger a_{n+1}\sigma^\dagger+a_n a_{n+1}^\dagger\sigma^{-})\nonumber\\
&+&\sum_{n=1,2}(E_n(t)a_n e^{i\omega_n t}+E_n^*(t)a_n^\dagger e^{-i\omega_n t}).
\label{Hamiltonian1B}
\end{eqnarray}
By applying the following subsequent unitary transformation $H_2=U_2 H_1 U_2^\dagger$ with $U_2=\exp[(g_1\sin\theta/\Delta) (a_1^\dagger\sigma^{-}-a_1\sigma^\dagger)]$ up to the second order, one can finally obtain the following Hamiltonian $H_2$ for the system in the dispersive regime
\begin{eqnarray}
H_2&=&{E_{\rm 01}\over 2}\sigma_z + \sum_n\omega_n a_n^\dagger a_n+{(g_1\sin\theta)^2\over 2\Delta}(2 a_1^\dagger a_1+1)\sigma_z \nonumber\\
&+&\sum_n{1\over{n(n+1)}}{{g_1 g_n g_{n+1}}\over \Delta\omega_1}\sin\theta\sin 2\theta\nonumber\\
&\times&(a_1^\dagger a_n^\dagger a_{n+1}+a_1 a_n a_{n+1}^\dagger)\sigma_z\nonumber\\
&+&\sum_{n=1,2}(E_n(t)a_n e^{i\omega_n t}+E_n^*(t)a_n^\dagger e^{-i\omega_n t}).
\label{Hamiltonian2A}
\end{eqnarray}

\section{}
We will start with the following Hamiltonian describing the Josephson junction coupled to the transmission line resonator\cite{transmon}
\begin{equation}
H=4E_c\left({\hat n}-N_g-\delta {\hat n}_g\right)^2-E_J \cos\phi
\label{Hamilt0}
\end{equation}
where $N_g$ and $\delta {\hat n}_g$ represent the dc and ac components, respectively, of gate charge imposed by the transmission line resonator.
At the absence of $\delta {\hat n}_g$, the exact eigenvalues and eigenfucntions of the above Hamitonian have been obtained as follows
\begin{equation}
H \Psi_m (\phi)=E_m \Psi_m (\phi)
\end{equation}
where $E_m$ depends on the gate charge $N_g$ and $\Psi_m (\phi)$ is given by Mathieu function.
The finite $\delta {\hat n}_g$ induces a coupling term between the qubit states and the microwave photon field, which can be written by
\begin{equation}
H_c=-\sum_i 2\beta e V^{(i)}_{\rm rms} (a_i+a_i^\dagger)\left({\hat n}-N_g\right)
\label{Coupling1}
\end{equation}
where $\beta=C_g/C_\Sigma$ and $V^{(i)}_{\rm rms}=\sqrt{\hbar\omega_i/2C_r}$.

Here we want to obtain the coupling Hamiltonian $H_c$ in the energy eigenbasis $\Psi_m$ as follows
\begin{equation}
H_c=-\sum_i 2\beta e V^{(i)}_{\rm rms} (a_i+a_i^\dagger)\sum_{m,m'} \vert m\rangle \langle m\vert {\hat n}-N_g \vert m'\rangle \langle m'\vert.
\label{Coupling2}
\end{equation}
One can explicitly show that the diagonal components are given by the following relation $\langle m\vert {\hat n}-N_g \vert m\rangle=-(1/8E_c)(\partial E_m/\partial N_g)$ and
the off-diagonal components $\langle m\vert {\hat n}-N_g \vert m'\rangle=((E_m-E_{m'})/8E_c)\langle m\vert {\partial\over \partial N_g}\vert m'\rangle$ for $m\ne m'$.
For the transmon with $E_J/E_c>>1$, the following approximation holds ${\partial\over \partial N_g}\Psi_m (\phi)\cong i\phi \Psi_m (\phi)$ and hence
the most dominant off-diagonal terms are given by the following formula\cite{transmon}
\begin{equation}
\langle m\vert {\hat n}-N_g \vert m+1\rangle\cong -i \sqrt{m+1\over 2}\left( E_J\over 8E_c\right)^{1/4}.
\end{equation}
Hence the Hamiltonian $H_c$ can be written by
\begin{eqnarray}
H_c&\cong& \hbar\sum_{m,i} g_{m m}^{(i)}\vert m\rangle \langle m\vert (a_i+a_i^\dagger)
 +i\hbar\sum_{m,i} g_{m m+1}^{(i)}\nonumber\\ &\times&(\vert m\rangle
\langle m+1\vert
-\vert m+1\rangle \langle m\vert)(a_i+a_i^\dagger)
\label{Coupling1}
\end{eqnarray}
where $g_{m m}^{(i)}=-2\beta e V^{(i)}_{\rm rms} \langle m\vert {\hat n}-N_g \vert m\rangle$ and $g_{m m+1}^{(i)}=-2\beta e V^{(i)}_{\rm rms} (E_J/8E_c)^{1/4}\sqrt{(m+1)/2}$.
The diagonal matrix elements are given by the following formula
\begin{eqnarray}
\langle m\vert {\hat n}&-&N_g \vert m\rangle=(-1)^{m+1}\sin{2\pi N_g}\nonumber\\
&\times&{\sqrt{2\pi}\,2^{4m+2}\over m!}\left({E_J\over 2E_c}\right)^{{m\over 2}+{3\over 4}} e^{-\sqrt{8E_J/E_c}}.
\end{eqnarray}
We note that the coefficients $g_{m m}^{(i)}$ are maximized at $N_g=1/4$.

In order to obtain the squeezing term, we have applied the following two unitary transformations successively: $U_1=\exp\left[-\sum_{m,i} (g_{m m}^{(i)}/\omega_i)(a_i-a_i^\dagger)\vert m\rangle \langle m\vert\right]$ and $U_2=\exp\left[i\sum_{m=0,1}\beta_m \left(a_1\vert m+1\rangle \langle m\vert + a_1^\dagger
\vert m\rangle \langle m+1\vert\right)\right]$ with $\beta_m=g_{m m+1}^{(1)}/\Delta_m$, $\Delta_m=\Omega_{mm+1}-\omega_1$, and $\Omega_{mm+1}=\Omega_{m+1}-\Omega_m$.
Here we have included the three lowest qubit states to take into account the weak anharmonicity of transmon and the following conditions are required to hold $|g_{m m+1}^{(1)}/\Delta_m|<<1$ for $m=0,1$.
Following a complicated but straightforward algebra, we have reproduced the following Hamiltonian for transmon\cite{transmon}
\begin{equation}
H_{tr}={\hbar\over 2}\Omega_{01}^\prime\sigma_z +\left( \hbar\omega^\prime+\hbar\chi\sigma_z\right)a_1^\dagger a_1
\label{HamiltonianTransmon}
\end{equation}
where $\Omega_{01}^\prime=\Omega_{01}+\chi_{01}$, $\omega^\prime=\omega_1-{1\over 2}\chi_{12}$, and $\chi=\chi_{01}-{1\over 2} \chi_{12}=-g_{01}^2 E_c/\Delta_0(\Delta_0-E_c)$.
The Hamiltonian for squeezing term is given by
\begin{equation}
H_{sq}=\left(-\beta_0\gamma_{01}\sigma_z+{1\over 2}\beta_1\gamma_{11}(1+\sigma_z)\right)
\left(a_1^\dagger a_1^\dagger a_2+a_1 a_1 a_2^\dagger\right)
\label{SqueezingTransmon}
\end{equation}
where $\beta_0\gamma_{01}=g_{01}^2(g_{00}-g_{11})/\sqrt{2}\omega_1\Delta_0$ and ${1\over 2}\beta_1\gamma_{11}=g_{01}^2(g_{11}-g_{22})/\sqrt{2}\omega_1(\Delta_0-E_c)$ with the superscript for $g_{mn}^{(i)}$ omitted and $i=1$. Here the term with the coefficient ${1\over 2}\beta_1\gamma_{11}$ stems from the virtual processes involving the third qubit state $\vert 2\rangle$.
Since the charge sensitivity $g_{m m}^{(i)}$ of the qubit states increases with $m$, ${1\over 2}\beta_1\gamma_{11}>>\beta_0\gamma_{01}$ for $E_J/E_c>>1$.


\end{appendix}


\end{document}